\documentclass[prl,aps,amsfonts,amssymb,twocolumn,floats,showpacs]{revtex4}
\usepackage{graphicx}

\begin{document}

\title{Quantum chaos in disordered graphene}

\author{I. Amanatidis and S.N. Evangelou\footnote{e-mail:sevagel@cc.uoi.gr}}
\affiliation{Department of Physics, University of Ioannina,
Ioannina 45110, Greece}

\begin{abstract}
\par
\medskip
We have studied numerically the statistics for electronic states
(level-spacings and participation ratios) from disordered graphene
of finite size, described by the aspect ratio $W/L$ and various
geometries, including finite or torroidal, chiral or achiral
carbon nanotubes. Quantum chaotic Wigner energy level-spacing
distribution is found for weak disorder, even infinitesimally
small disorder for wide and short samples ($W/L>>1$), while for
strong disorder Anderson localization with Poisson
level-statistics always sets in. Although pure graphene near the
Dirac point corresponds to integrable ballistic statistics chaotic
diffusive  behavior is more common for realistic samples.

\end{abstract}

\pacs {73.22.Dj,73.63.-b,73.20.Fz,81.05.Uw}

\maketitle
\section{I. INTRODUCTION}

\par
\medskip
During the last few years theoretical and experimental interest
has grown immensely  on monolayer graphite samples, known as
graphene following its pioneering fabrication \cite{r1,r2,r3}.
This remarkable two-dimensional (2D) honeycomb lattice structure
made of carbon atoms behaves rather differently from ordinary 2D
metals. Its most interesting quantum property is a linear
dispersion near the Fermi energy, known as the Dirac point, where
the conduction and valence bands of graphene touch each other,
with the low-lying excitations behaving as massless Dirac
fermions. Anomalous integer quantum Hall effect for graphene has
been observed  \cite{r2,r4}, which depends on the symmetry of
introduced disorder \cite{r5} and has been discussed in terms of
relativistic Dirac theory\cite{r6}. Due to its extraordinary
properties, such as high mobility, current-carrying capacity and
thermal conduction, graphene is regarded as a good candidate to
replace silicon in future nanoelectronics \cite{r7}. The
possibility of  gating and processing graphene sheets into
multi-terminal devices is also an intriguing issue for related
research \cite{r8}.

\par
\medskip
The purpose of this paper is to examine the energy
level-statistics for finite graphene sheets (quantum dots in the
form of nanotubes) \cite{r9}, for width $W$ and length $L$
characterized by their aspect ratio $W/L$. We have also considered
various directions of graphene and various sorts of boundary
conditions (BC), in the presence of disorder. Although graphene is
widely regarded as a rather clean material it is believed that
mild disorder should exist \cite{r10,r11,r12,r13,r14,r15}. For
example, alloy type disorder from placing at random different
lattice atoms or vacancies in the honeycomb lattice \cite{r13}, or
correlated disorder \cite{r14}. The latter, by changing the
universality class from orthogonal to symplectic \cite{r14}, is
expected to lead to unusual phenomena, such as weak
anti-localization with metallic behavior at the Dirac point. The
usual short-ranged disorder causes localization also in graphene
\cite{r15}, which can limit the performance of graphene made
devices. These effects of disorder naturally reflect on the
conduction of electrons, which can be studied by usual scattering
methods for electron waves in graphene billiards \cite{r16}. Near
the Dirac point, where the density of states for pure graphene
vanishes, the transmission becomes pseudo-diffusive in the limit
($W/L>>1$) independently of BC \cite{r17}.

\par
\medskip
Recent experimental observations for size quantization of electron
levels in graphene quantum dots \cite{r18} led to theoretical
consideration of level-statistics in tight-binding
models\cite{r19}. These numerical simulations for special
geometries of the honeycomb lattice (named Dirac billiards, with
geometrical shapes such as weakly disordered circle or triangle)
are in reasonable agreement with the observed level-statistics in
the experiment \cite{r18}. It was found that weak disorder at the
edges of the sample can also produce quantum chaos, which is
independent of the shape of graphene dots. Our results  agree with
these observations \cite{r18,r19} since we have also obtained
Poisson (integrable) level-statistics in the clean limit, which
becomes chaotic for weak disorder, independently of geometry.
Moreover, our study which was carried out for all strengths of
disorder finds quantum chaotic behavior for wide and short
nanotubes ($W/L>>1$) even in the infinitesimally small disorder
limit, confirming the presence of pseudo-diffusive behavior in
this case. In the opposite limit of narrow long cylinders
($W/L<<1$) the statistics becomes ballistic $\delta$-function
type, similar to that of the one-dimensional (1D) chain. In other
words, for realistic samples having arbitrary small disorder the
outstanding ballistic behavior of long nanotubes is replaced by
diffusive chaotic behavior for short and wide graphene samples.

\par
\medskip
The paper is organized as follows: In Sec. II we introduce the
graphene lattices studied. They are of various sizes and
geometries, corresponding to chiral and achiral carbon nanotubes.
Their tight-binding Hamiltonian in the presence of disorder is
introduced. In Sec. III we show that for diagonal and off-diagonal
disorder (in the case of large $W/L$ even for infinitesimally
small disorder) Wigner chaotic level-statistics is obtained while
for strong disorder it becomes Poisson, independently of the type
of graphene sheet. Our results for the participation ratio
complement these findings by showing extended eigenstates close to
the Dirac point and confirm the presence of Anderson localization
for strong disorder. Finally, in Sec. IV we summarize our main
conclusions.

\par
\medskip
\section{II. The tight binding Hamiltonian}

\par
\medskip
The tight-binding Hamiltonian of $\pi$ bonding in graphene is
 \begin{equation}
      H=\sum_{i}\varepsilon_{i}c_{i}^{\dag}c_{i} -\gamma
      \sum_{<i,j>}(c_{i}^{\dag}c_{j}+c_{j}^{\dag}c_{i}),
\end{equation}
where $c_{i} (c_{i}^{\dag})$ annihilates(creates) an electron at
the sites of the honeycomb lattice, $\varepsilon_{i}$ represents
the on site energy which in the case of diagonal disorder is a
random number in the range $[-w/2,w/2]$, where $w$ denotes the
disorder. For off-diagonal disorder $\varepsilon_{i}=0$ and the
logarithm of the nearest neighbor hopping element $ \gamma$ takes
random values in the range $[-w/2,w/2]$ \cite{r20}.

\par
\medskip
In the case of periodic BC only in one direction, the other being
much longer with hard wall BC, the graphene samples represent
finite single wall carbon nanotubes with $W/L<<1$ \cite{r9}. These
are effectively one-dimensional objects, build up by wrapping
sheets of graphene into very long cylinders, and in modern
technology underpin the evolution to nano-electronics. Their
electronic properties depend on the wrapping direction which is
characterized by the so-called chiral vector (n,m), expressed in
terms of graphene primitive unit lattice vectors
$\overrightarrow{a}_1$, $\overrightarrow{a}_2$,
$|\overrightarrow{a}_1|=
|\overrightarrow{a}_1|=\sqrt{3}a_{C-C}=a$, where $a_{C-C}$ is the
distance between two atoms and $a$ the honeycomb lattice constant.
Since various possibilities exist to roll up graphene, depending
on the rolling direction one obtains armchair, zig-zag and chiral
nanotubes \cite{r9}.  The  armchair nanotubes characterized by a
chiral vector (n,n) can be obtained by rolling the graphene sheet
along one of the three vectors joining a honeycomb lattice site to
its nearest-neighbors. The perimeter of the (n,n) armchair
nanotubes consists of n hexagons along the rolling direction
connected by n single bonds and they are always metallic.  The
(n,0) zig-zag nanotubes are defined by a rotation of the graphene
sheet by 90$^{0}$ and the n hexagons followed by n single bonds
lie now along the longitudinal axis of the tube (the rolling
direction is perpendicular). The  (n,0)  nanotubes are metallic
only when $n$ is a multiple of 3. In all other cases, that is
rotating the graphene sheet in between the previous two and then
rolling it up, is equivalent to placing the hexagons followed by
single bonds along intermediate directions. This third type of
carbon nanotubes is called chiral (n,m), $n\neq m$ and its band
structure is metallic only when $n-m$ is a multiple of 3
\cite{r9}.

\par
\medskip
Our approach consists of studying the effects of disorder in
graphene by obtaining numerically the eigenvalues and eigenvectors
of the Hamiltonian $H$ in the presence of disorder. In our
computations we have fixed the short-ranged site (diagonal) and
bond (off-diagonal) type disorder for many random samples.  The
statistics of the obtained eigen-solutions are subsequently
analyzed in order to address the quantum chaos issues.  Since for
the $(n,n)$ armchair nanotube a unit cell along its long direction
consists of two slices around the tube, (with $2n$ carbon atoms
each making $4n$ atoms in total the total number of hexagons
covering the unit cell is $N_{hex}=2n$). The number of such unit
cells along a finite nanotube is $N_c$ so that their size can be
classified in terms of $n$ and $N_c$.

\par
\medskip
In the absence of disorder ($w=0$) the dispersion of pure graphene
sheet taken in the correct orientation, that of armchair nanotubes
along the $x$-axis with their rolling direction along the vertical
$y$-axis, is \cite{r21}
\begin{eqnarray}
 E(k_x,k_y) = \pm \gamma \left(1+4\cos(\frac{\sqrt{3}k_ya}{2})\cos({\frac {k_xa}{2}})
 \right.
  \nonumber\\
 \left. +4\cos^{2}({\frac {k_xa}{2}})\right) ^{1/2},
\end{eqnarray}
with quantized values of ${\bf{k}}=(k_x,k_y)$. For a finite
armchair nanotube (periodic BC along $y$ and hard wall BC along
$x$) the eigenstates are labelled by two integers, via
\begin{equation}
      k_x a={\frac {\pi}{N_c+{\frac{1}{2}}}}j_x, ~~~k_y a={\frac {2\pi}{\sqrt{3}n}}j_y,
\end{equation}
with $j_x=1,...,N_c$ and $j_y=1,...,2n$. The output is $2nN_{c}$
positive and exactly equal negative eigenvalues (a fact of
sublattice symmetry \cite{r22}, since the two types of atoms A and
B in graphene for nearest neighbor hopping make two interconnected
A and B sublatices). In the case of infinite nanotubes the $k_x a$
quantization of Eq. (3) does not arise and the band structure of
Eq. (2) becomes one-dimensional, a function of $k_x$ only
\cite{r9}. For toroidal armchair nanotubes, where appropriate
periodic BC are imposed also along the longitudinal direction $k_x
a={\frac {2\pi}{N_c}}j_x$, $j_x=1,...,N_c$.

\par
\medskip
For the more general chiral (n,m), $n\neq m$ nanotubes the
quantized component of the wave vector (n,m) along the perimeter
of the nanotube where periodic BC are imposed, satisfies
\begin{equation}
     k_ya=\frac{2\pi }{W/a}j_y, ~~j_y=1,....N_{hex},
\end{equation}
having width $W=a\sqrt{n^2 + m^2 + nm}$ given by their perimeter
and $N_{hex}=2L^2/(a^2d_r)$ hexagons  in a unit cell  of length
$\mathbf{|T|}=\sqrt{3}W/d_r$, $d_r$ being the greatest common
divisor of 2m+n and 2n+m. For a given value of the quantized $k_y$
the energy of an infinite chiral nanotube is a function of the
continuous longitudinal component $k_x$ only, with
$-\frac{\pi}{\mathbf{|T|}}<k_x<+\frac{\pi}{\mathbf{|T|}}$. For
armchair nanotubes (n,n) the unit cell reduces to width
$W=a\sqrt{3}n$, length ${\mathbf{|T|}}=a$ and $N_{hex}=2n$. Since
$L=N_{c}a$ for $N_c$ unit cells the considered aspect ratio is
$W/L=d_r/(\sqrt{3}N_c)$ for chiral (n,m) nanotubes, reducing to
$W/L=\sqrt{3}n/N_c$ for armchair (n,n) nanotubes.

\begin{figure}
\includegraphics[angle=0,width=4.0cm]{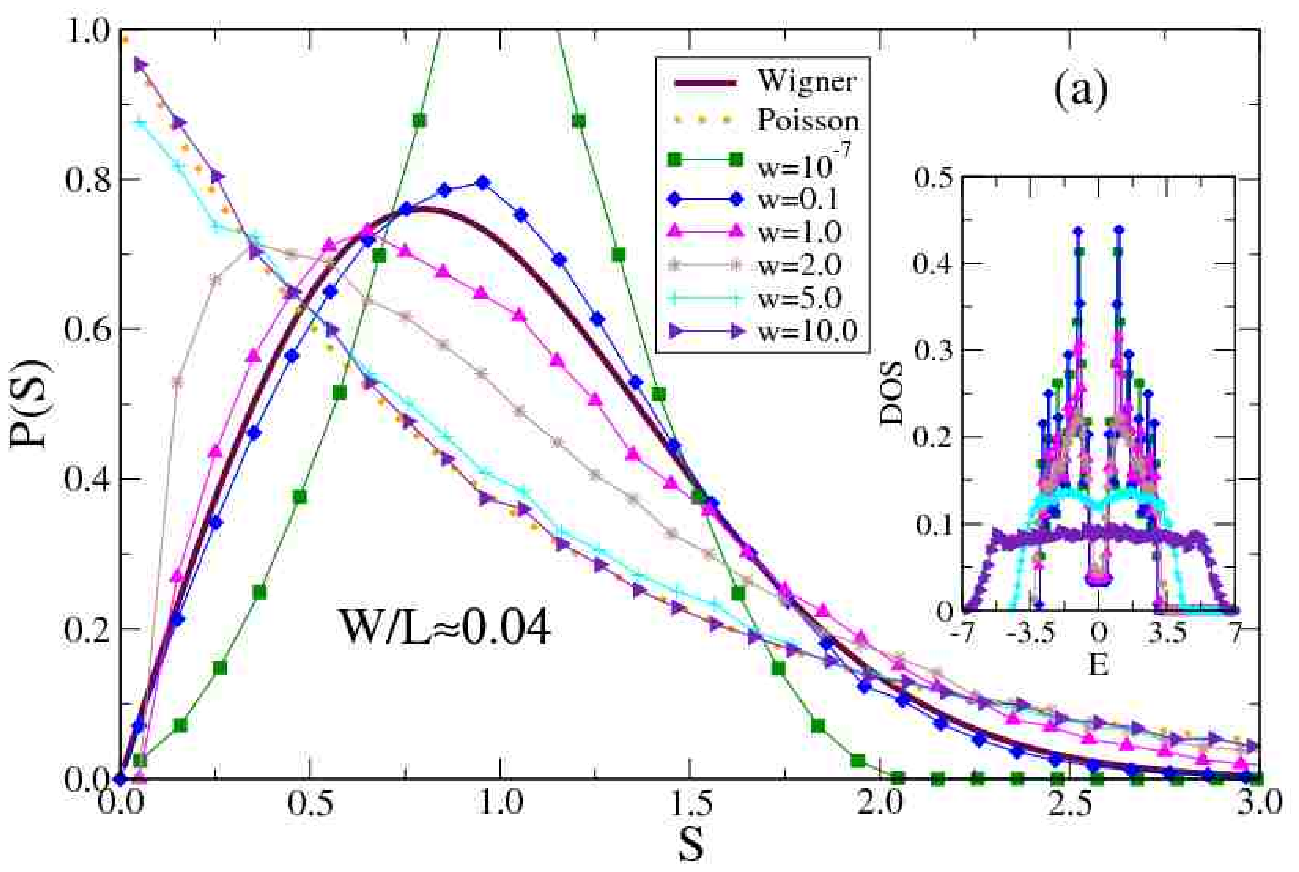}
\includegraphics[angle=0,width=4.0cm]{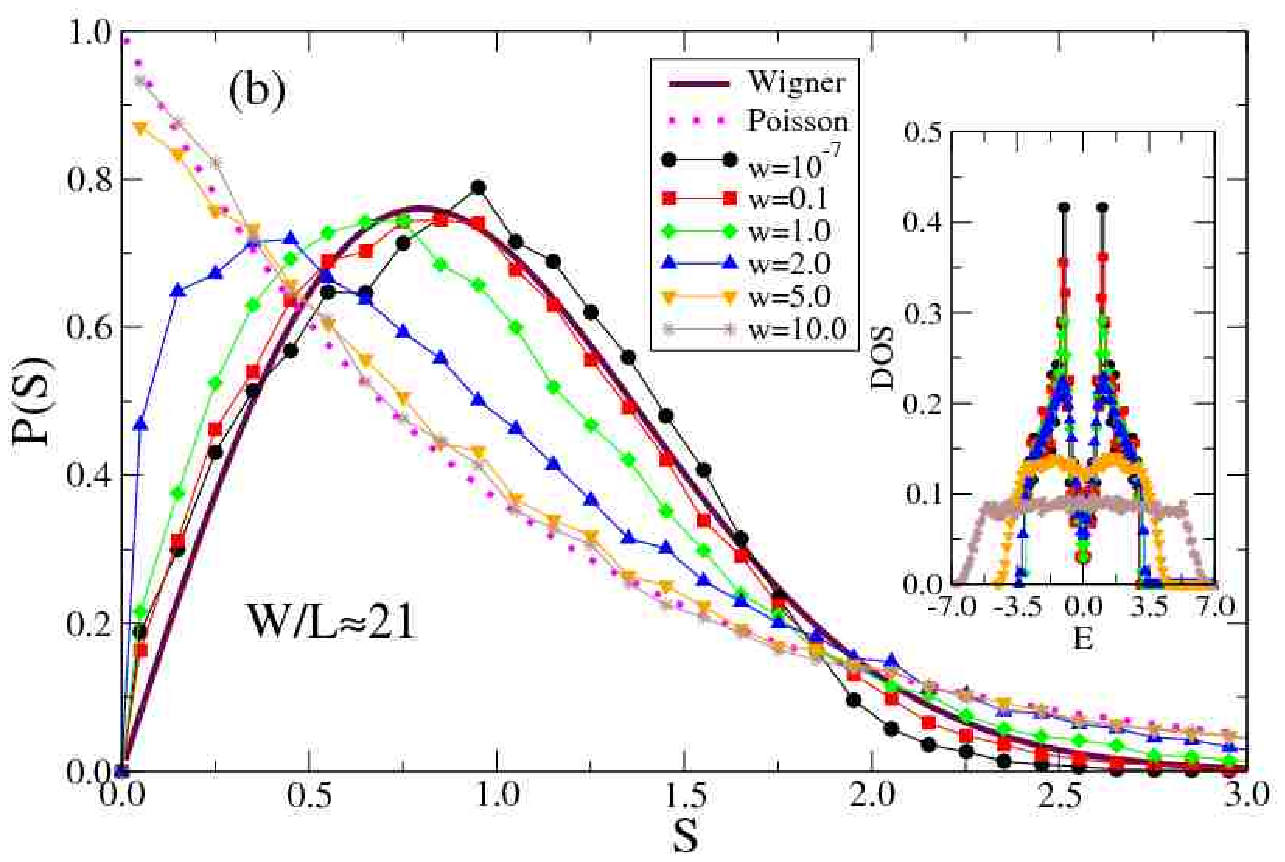}
\includegraphics[angle=0,width=4.0cm]{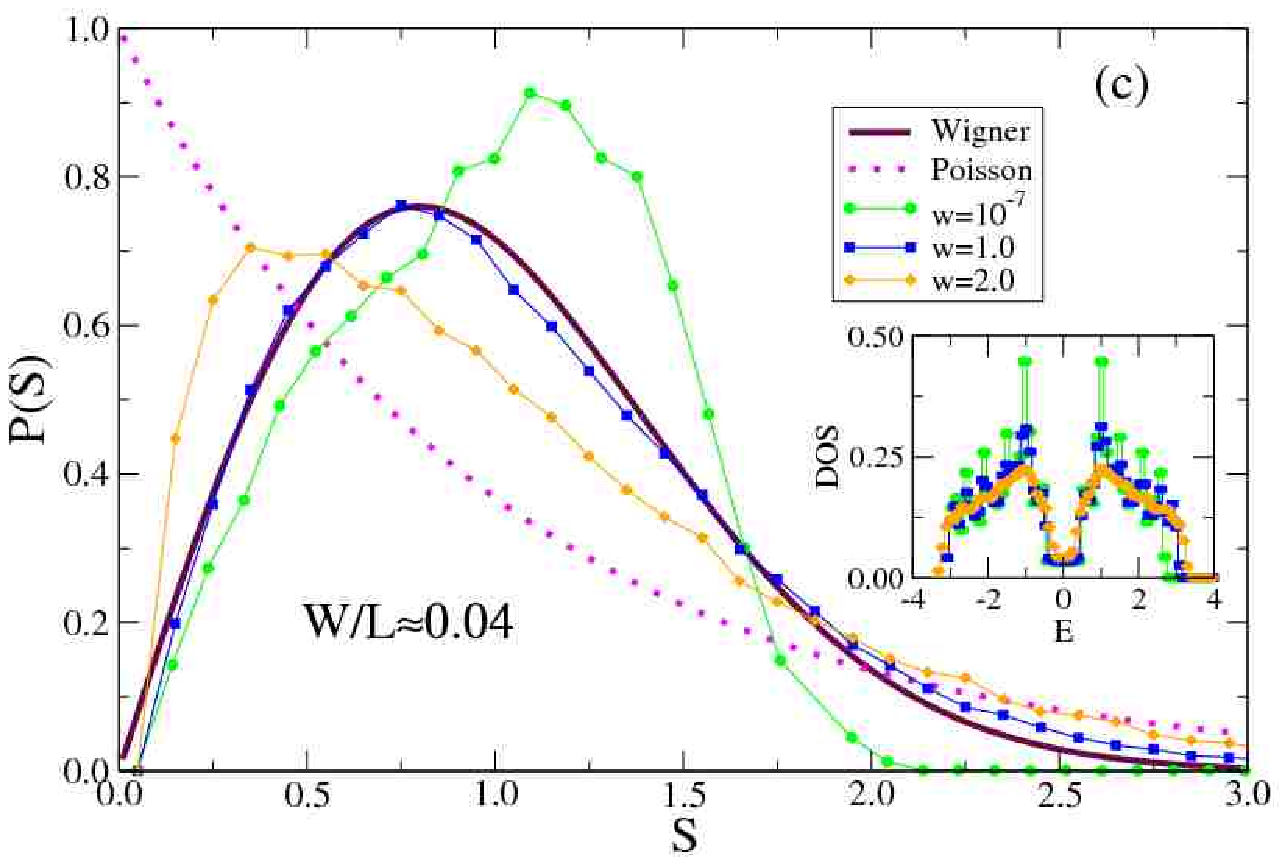}
\includegraphics[angle=0,width=4.0cm]{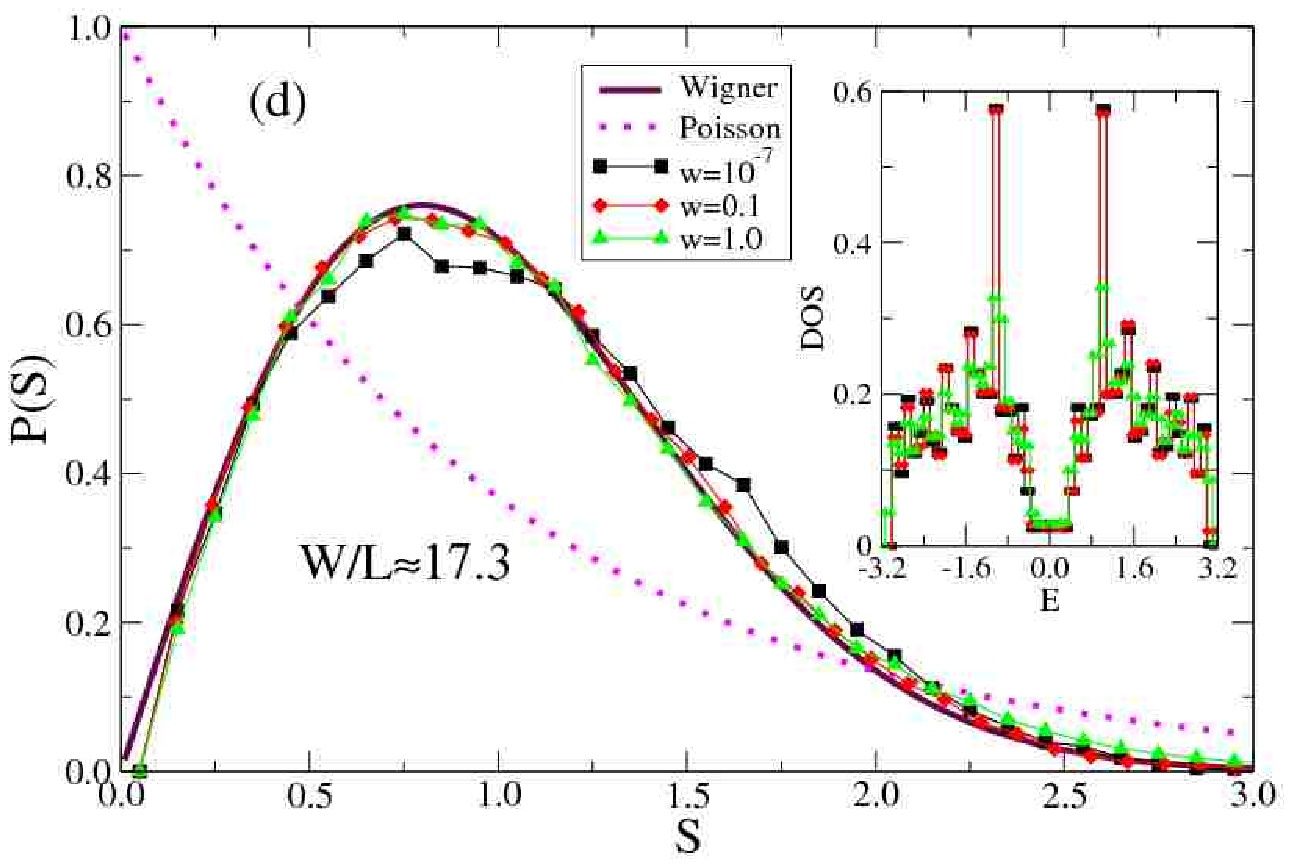}
\caption{The normalized
level-spacing distribution function $P(S)$ from 10 samples of graphene with
diagonal disorder (for the values of disorder strength $w$ see figure): {\bf (a)}
For graphene structures with a very small aspect ratio $W/L\approx 0.04$ and
$4960$ sites corresponding to $N_c=40$ unit cells of the chiral (7,4) nanotube.
As the disorder strength increases a gradual crossover is seen, from a broad
$\delta$-function (in the form of a Gaussian around the mean) for clean nanotubes
to a Poisson distribution for strong disorder. {\bf (b)} The same as in (a) but
for a wide and short chiral (36,180) nanotube with $N_{c}=1$ having aspect ratio
$W/L\approx21$ ($4464$ sites). In this case the distribution is close to Wigner
even for almost zero disorder. {\bf (c),(d)} From the (7,4) and (36,120) toroidal
nanotube geometries consisting of $N_c=40,12$ unit cells corresponding to aspect
ratios $W/L\approx0.04, 17.3$ and sites $4960, 5760$ sites, respectively. In the
insets the densities of states are shown with the characteristic dip near the
Dirac point, which remains for weak disorder but closes for strong disorder
($w>3$), leading to a broad density of localized states.}
\end{figure}

\begin{figure}
\includegraphics[angle=0,width=4.0cm]{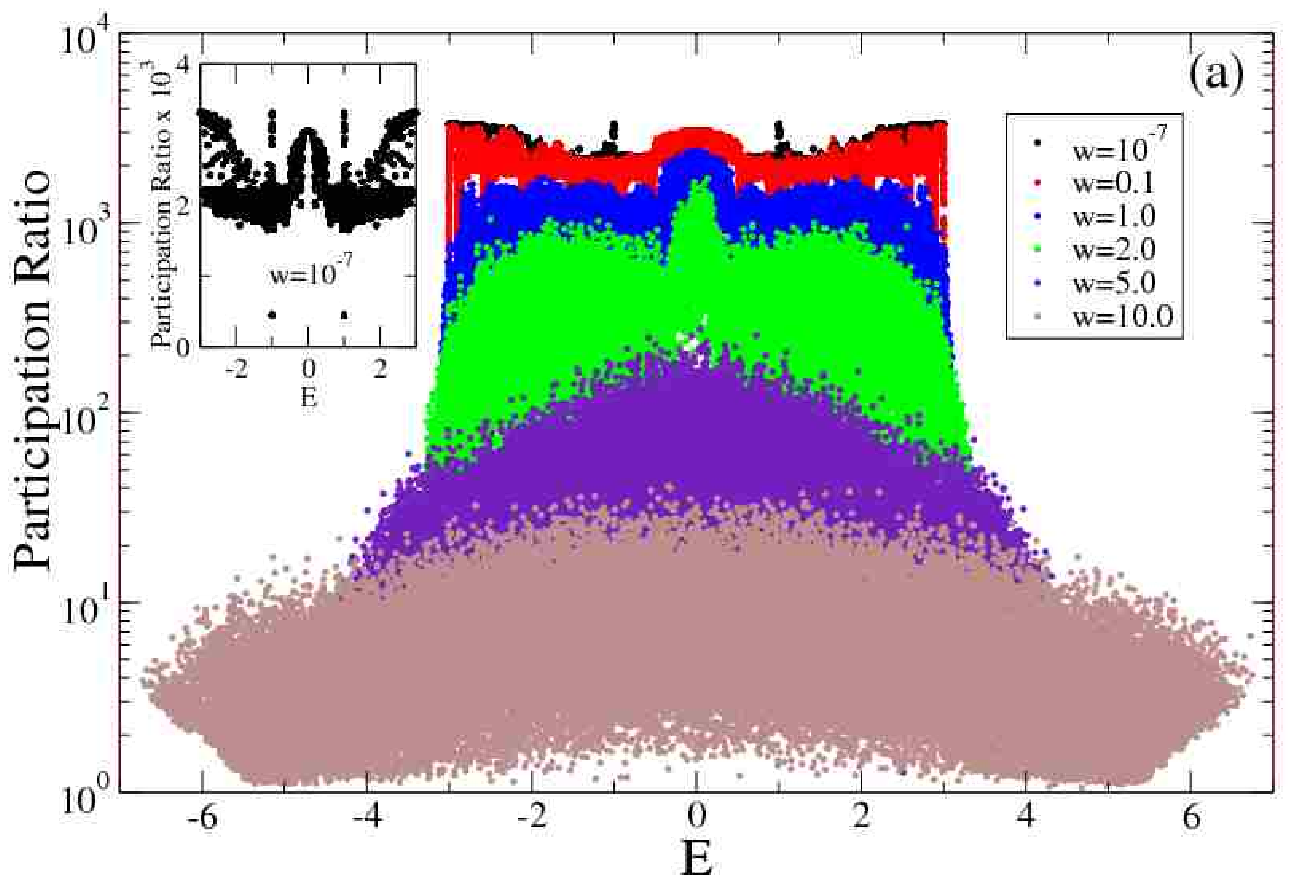}
\includegraphics[angle=0,width=4.0cm]{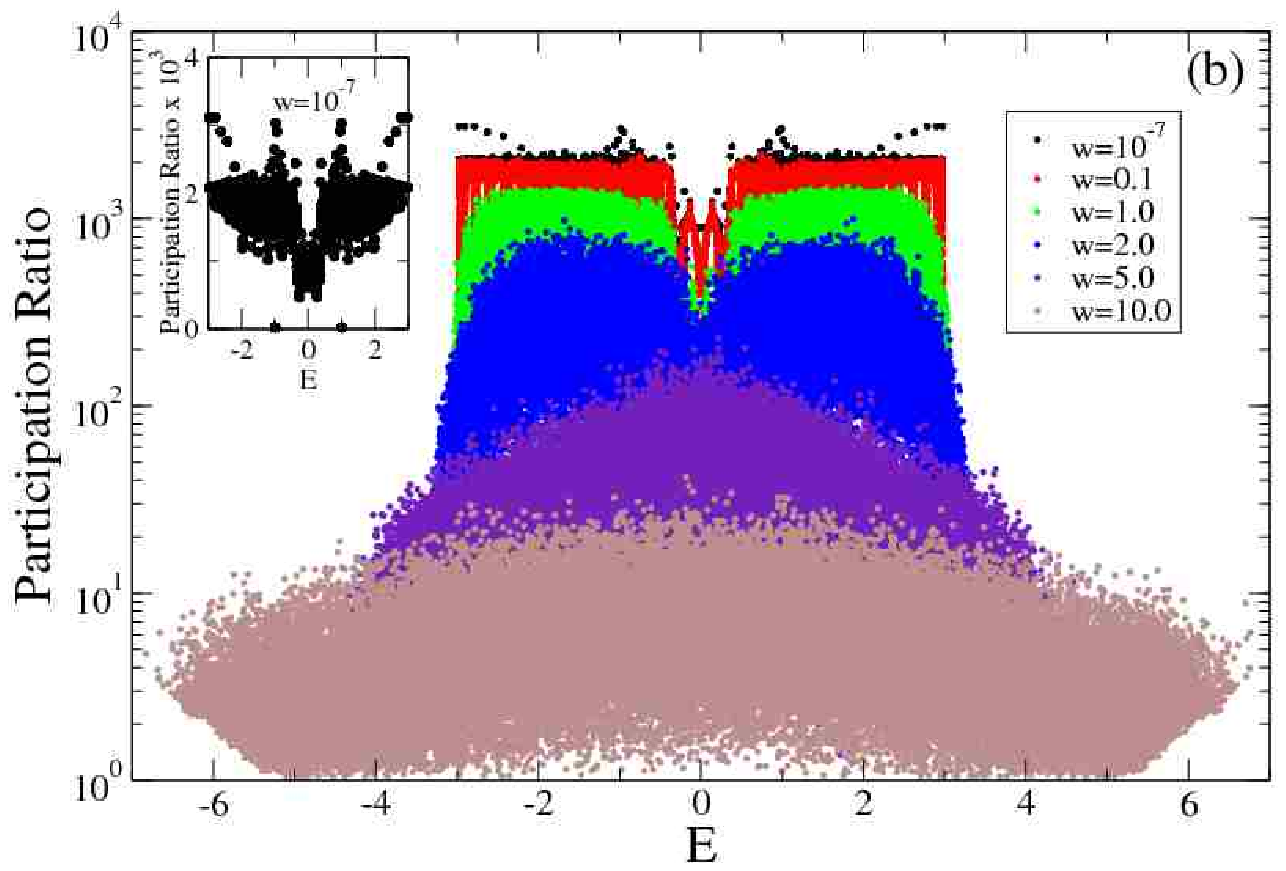}
\includegraphics[angle=0,width=4.0cm]{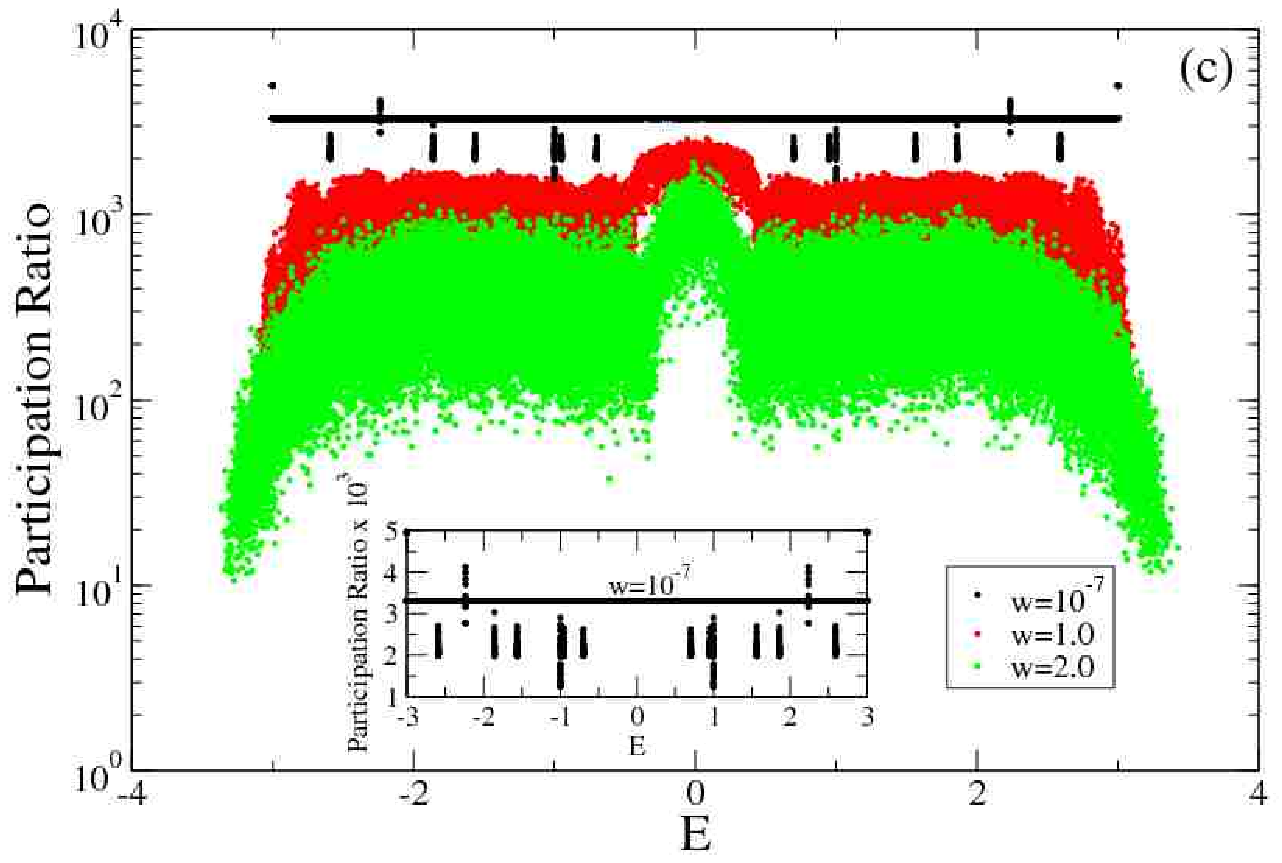}
\includegraphics[angle=0,width=4.0cm]{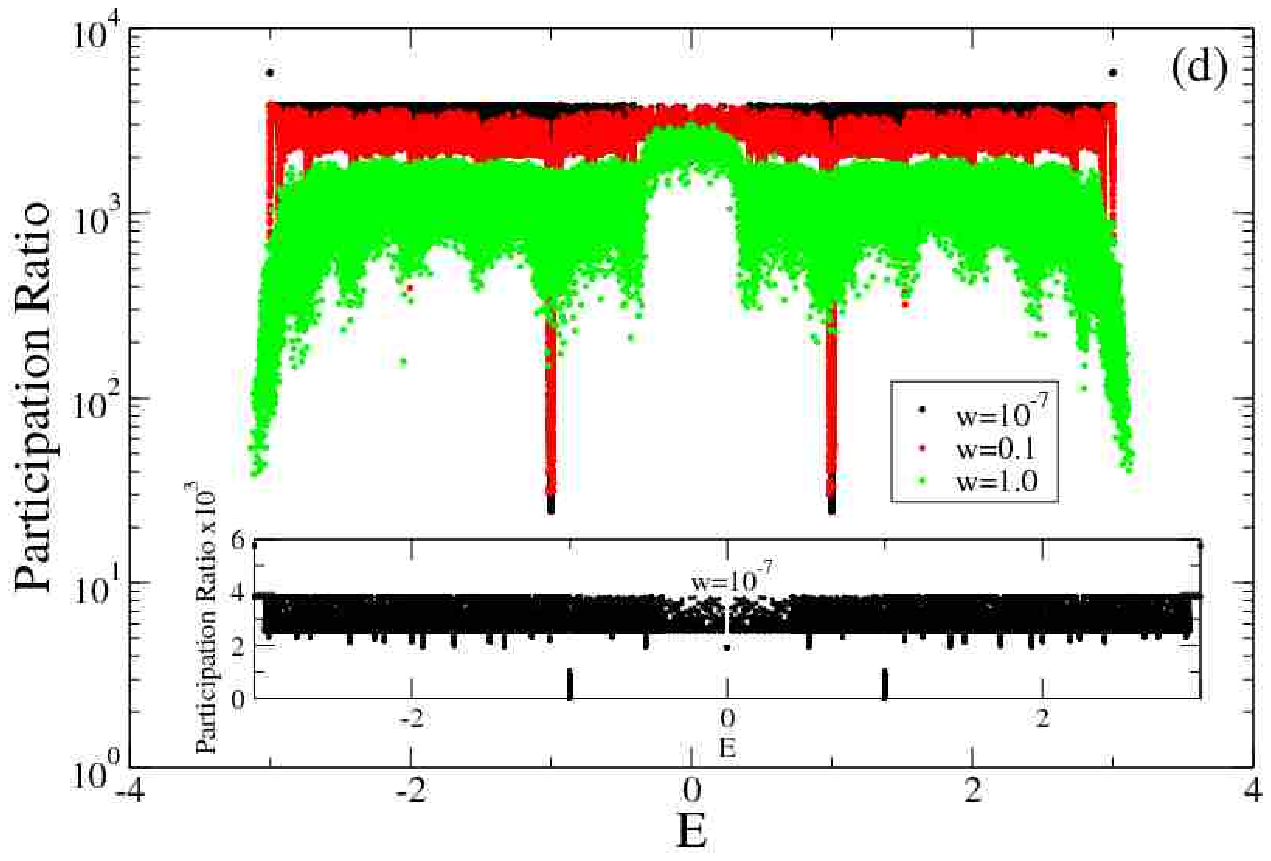}
\caption{For the same parameters as in Fig.1, but for the
participation ratios of the corresponding eigenstates  versus
energy. The participation ratios indicate the number of sites
where each eigenstate has a significant amplitude. For a perfect
extended state its value should be $\approx 5000$ sites for the
chosen sizes. In the insets the results for pure graphene $w=0$
are magnified, note the linear scale on the $y$-axis.}
\end{figure}

\par
\medskip
\section{III. Results}

\par
\medskip
\subsection{Pure graphene sheets}

\par
\medskip
Numerical computations of the eigen-solutions are required for
chiral nanotubes even in the absence of disorder, since their unit
cell along the tube can become arbitrarily large depending on
(n,m). For pure graphene we have found the expected ballictic
behavior manifested via integrable Poisson statistics for
arbitrary $W/L$, which becomes ballistic 1D-like in the limit
$W/L<<1$.  This can be simply understood near the Dirac point
where the density of states is linear. In this regime the
integrated density of states is ${\cal N}(E)\propto E^{2}$.
Therefore,  the statistics of the energy levels $E$ should be
replaced by the statistics of the unfolded squared levels $E^{2}$
which is required in order to make the density of states constant
by fixing the mean level-spacing. If we use the usual quadratic
dispersion for the squared levels $E^{2} \propto k_{x}^{2}+k_y^
{2}$  the statistic  near the Dirac point becomes equivalent to
the statistics of integrable billiards \cite{r23}, with $E^{2}$
replaced by $E$ from
\begin{equation}
E_{j_x,j_y}=\alpha j_{x}^{2}+j_{y}^{2}, ~~~j_x,~j_y~integers,
\end{equation}
$\alpha$ being an irrational number fixed by the size of the
adopted sample. The resulting statistics for zero disorder is
obviously Poisson \cite{r23}, except for very long samples in the
1D limit where the sum of Eq. (5) should depend on one parameter
only and the corresponding level statistics becomes a trivial
$\delta$-function. We have verified this behavior, in agreement
with \cite{r19}. However, for infinitesimally small disorder as
seen in Fig. 1(b) the statistics changes and becomes chaotic when
$W/L$ is large (Wigner distribution) while it remains trivial
ballistic $1D$-like for small $W/L$  shown in Fig. 1(a) (broad
$\delta$-function). In the almost clean limit, the ballistic
behavior for small $W/L$ is replaced by chaotic diffusive for
larger $W/L$ samples.

\par
\medskip
In the insets of Fig. 2 the behavior of the corresponding
eigenstates for pure graphene can be seen on a linear scale, via
their participation ratios. The main observation is that for the
small $W/L$ (Fig. 2(a) inset) the majority of ballistic states
have a higher participation ratios when compared to those with
large $W/L$ (Fig. 2(b) inset). The distribution of the
participation values becomes sharper for the small $W/L$ torroidal
nanotubes (Fig. 2(c) inset) when compared to large $W/L$ (2(d)
inset).

\par
\medskip
\subsection{Diagonal and off-diagonal disorder}

\par
\medskip
The density of states and the level-spacing distribution function
$P(S)$ are computed for various graphene sheets with aspect ratios
$W/L$, including chiral nanotubes. The corresponding nearest-level
spacing distribution function is illustrated for different values
of diagonal disorder in Fig. 1. All the unfolded energy levels in
the band are considered for $10$ runs, making the density of
states constant everywhere in the band. No significant difference
was found when levels near the Dirac point  are considered only.
For small $W/L$ (Fig. 1(a)) the level spacing distribution
function $P(S)$ is shown to vary from a broad $\delta$-function
like curve for almost zero disorder ($w=10^{-7}$)  towards the
Poisson localized limit for strong disorder $w>>$. For large $W/L$
we observe instead (Figs. 1(b),(d)) a crossover not from a
$\delta$-function but from a Wigner chaotic distribution towards
the localized Poisson limit in the strong disorder limit($w>>$).
In the insets of Fig. 1 the corresponding average density of
states is shown with the dip near the Dirac point, which
disappears when the disorder increases.

\par
\medskip
In Fig. 2 we demonstrate the participation ratio which  measures
the extend of the corresponding eigenstates  on the lattice. Near
the Dirac point the values are seen to be higher for small $W/L$
(Fig. 2(a)) when compared to large $W/L$ (Fig. 2(b)). This
observation is related to the observed $\delta$-function and
Wigner $P(S)$ distributions in Fig. 1, for small and large $W/L$,
respectively. Similar results are obtained for torroidal nanotubes
(Fig. 2(c) and (d)). For strong disorder our results indicate
localization of all the states in the band with their
participation ratio becoming small.

\par
\medskip
Off-diagonal disorder preserves chiral symmetry since it connects
one sublattice to the other \cite{r20}. In Fig. 3 we show our
results for the chiral (7,4) nanotube with off-diagonal disorder
in the small $W<<L$ nanotube limit. Fig 3(a) shows $P(S)$ (with
the density of states in the inset) and fig. 3(b) the
participation ratio of the corresponding eigenstates. Similar
results are obtained to the diagonal disorder case, with the
exception of the appearance of a singularity in the density of
states, which develops at the band center (not seen in the figure
due to scale) \cite{r20}.

\begin{figure}
\includegraphics[angle=0,width=7.0cm]{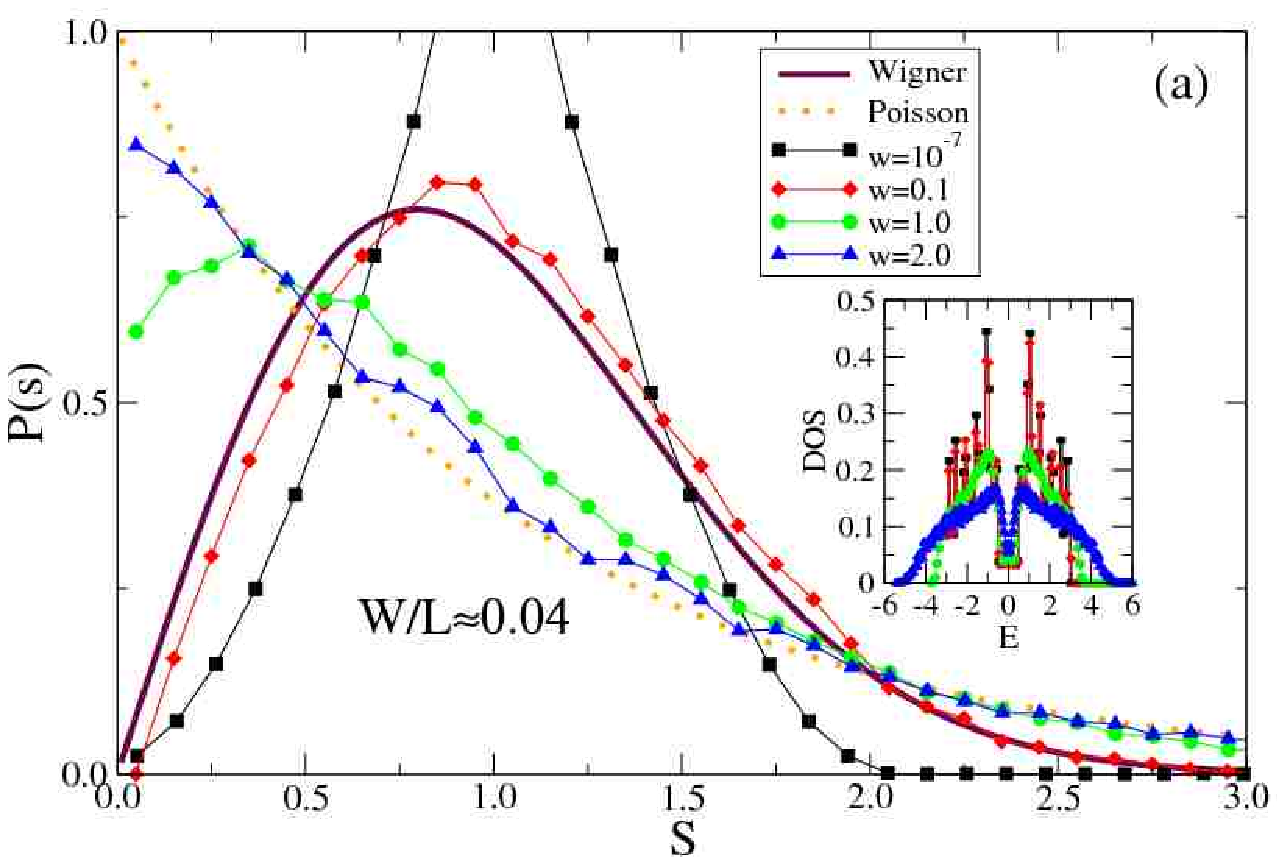}
\includegraphics[angle=0,width=7.0cm]{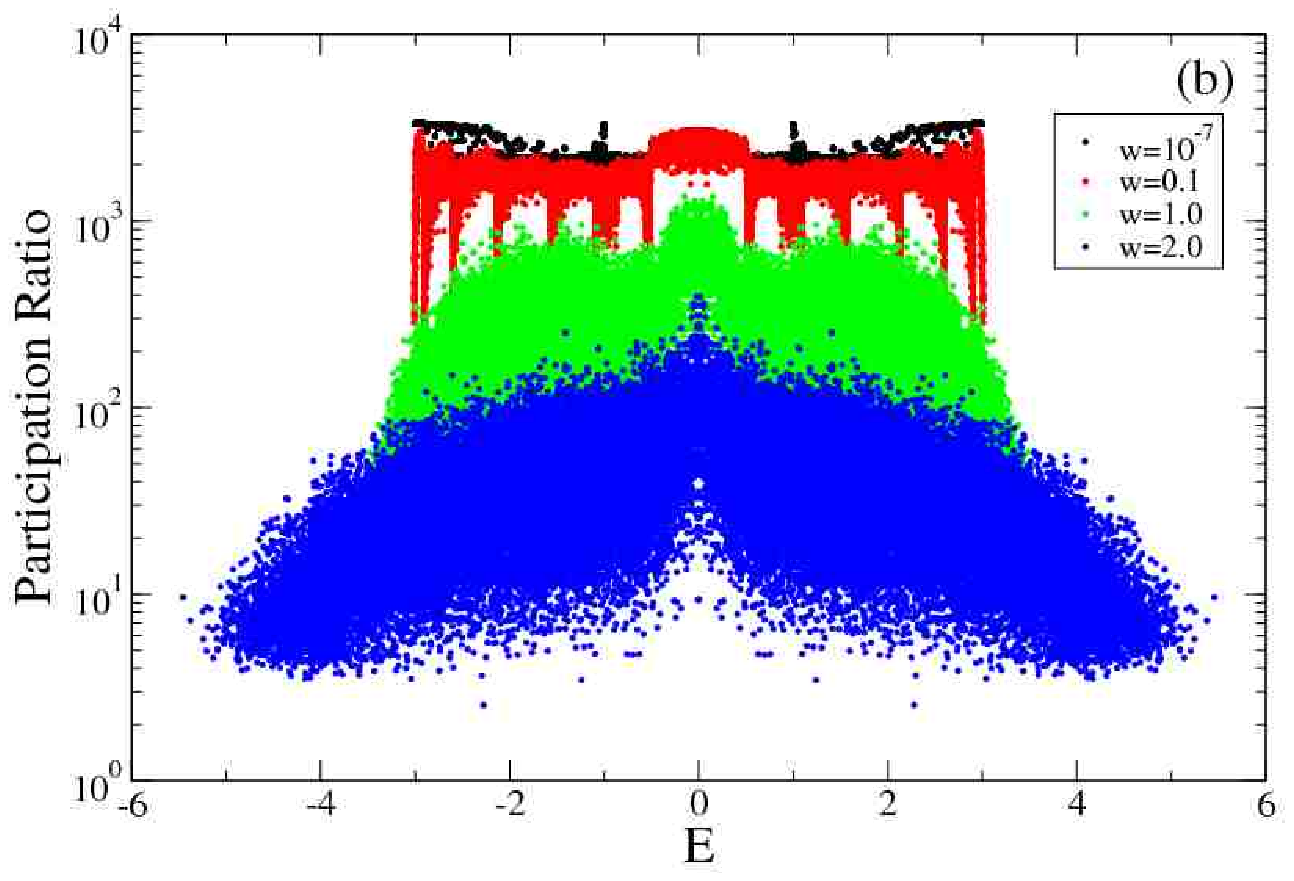}
\caption{{\bf (a)} The level-spacing distribution function $P(S)$
and {\bf (b)}the participation ratio for off-diagonal disorder of
strength $w$ (see figure) for the small aspect ratio
$W/L\approx0.04$ chiral (7,4) nanotube geometry of $N_c=40$ cells
($4960$ sites).}
\end{figure}


\par
\medskip
\section{IV. DISCUSSION}

\par
\medskip
It is known that the band structure of graphene is metallic
without a gap, having a density of states which becomes zero at
the Dirac point. However, depending on the geometry of the samples
a small gap can open near the Dirac point, e.g. for the (n,m)
nanotubes which become insulating when $n-m$ is not a multiple of
3. Our study of the eigensolution statistics for various
geometries specified by the aspect ratio $W/L$ in the presence of
disorder reveals some interesting features: First, quantum chaos
becomes relevant also for weakly disordered graphene, with
pseudo-diffusive Wigner statistics even for almost zero disorder
when $W/L$ is large. Second, for strong disorder localization
occurs for all states in the band, in agreement with finite-size
scaling transfer matrix studies at the Dirac point. Third, near
the Dirac point, due to the linearity of density of states of pure
graphene, a natural unfolding of levels corresponds to the study
of squares of energies $E^{2}$ which gives integrable irrational
billiard statistics. However, the pseudo-diffusive quantum chaotic
behavior of graphene for infinitesimally small disorder
independently of BC, for large aspect ratio $W/L$ should become
obvious in scattering experiments from nanotubes.

\par
\medskip
Since infinitesimal disorder is inevitable also for graphene the
question whether ballistic or diffusive behavior prevails can be
answered in favor of a simple 1D-like ballistic behavior for small
$W/L$ and pseudo-diffusive for large $W/L$. For weak disorder
quantum chaos and for strong disorder localization occurs, in
agreement with other 2D disordered lattices. In conclusion, our
computations for graphene in the presence of disorder show quantum
chaos or Anderson localization for weak disorder (even almost zero
disorder for large $W/L$) or strong disorder, respectively. This
demonstration of quantum chaos in graphene is in agreement with
recent theoretical and experimental studies.

\end{document}